\documentclass[a4paper,12pt]{article}

\usepackage{amsmath}
\usepackage[dvips]{geometry}
\usepackage[T1]{fontenc}
\usepackage{graphicx}
\usepackage{url}
\usepackage{citesort}
\usepackage[amssymb]{SIunits}

\newcommand{\J}{\mathrm{j}}  
\newcommand{\D}{\mathrm{d}}  
\newcommand{\E}{\mathrm{e}}  

\newcommand{\bemlib}{\textsc{bem3d}}
\newcommand{\wmpi}{\textsc{wmpi}}
\newcommand{\sisl}{\textsc{sisl}}
\newcommand{\gmc}{\textsc{gmc}}

\newcommand{\toprule}{\hline}
\newcommand{\midrule}{\hline}
\newcommand{\bottomrule}{\hline}

\begin{document}

\bibliographystyle{unsrt}



\title{BEM3D: a free adaptive fast multipole boundary element library}

\author{M.~J.~Carley}

\maketitle






\begin{abstract}
  The design, implementation and analysis of a free library for
  boundary element calculations is presented. The library is free in
  the sense of the GNU General Public Licence and is intended to allow
  users to solve a wide range of problems using the boundary element
  method. The library incorporates a fast multipole method which is
  tailored to boundary elements of order higher than zero, taking
  account of the finite extent of the elements in the generation of
  the domain tree. The method is tested on a sphere and on a cube, to
  test its ability to handle sharp edges, and is found to be accurate,
  efficient and convergent.
\end{abstract}


\section{INTRODUCTION}
\label{sec:intro}

The boundary element method has become accepted as a standard
computational technique for many problems and, when combined with a
fast multipole method, it has proven capable of solving very large
problems, primarily in scattering calculations for acoustic and
electromagnetic applications. This paper describes \bemlib, a free
library released under the GNU General Public Licence, for the
solution of general problems using the boundary element method,
incorporating a fast multipole method which is tailored to boundary
element problems using higher-order elements.

Free software has become an important part of the resources available
to those working in scientific and computational fields over the last
few years. Standard codes and libraries for many of the operations
routinely carried out in computation are now freely available and,
because their source code is openly published, they can be modified
and improved by experts in the field. Because codes are distributed as
libraries for a specific purpose, it is possible to select modules
which can be combined in codes for particular applications. This
avoids the problem of reinventing the wheel, speeding code development
and improving reliability. 

The library described in this paper makes use of a number of existing
free libraries~\cite{popinet00,galassi-davies-etal05,lapack99} for
computational applications, as well as a standard library of
portability and utility functions. The existing software is extended
by adding functions for the handling of boundary elements, for fast
multipole calculations and some support for parallel systems. The
intention is that other users will be able to make use of \bemlib\ to
solve their own problems by adding a Green's function for the partial
differential equation in question. This paper describes the design
choices which have been made in \bemlib\ and the implementation of
certain features including a novel adaptive fast multipole method for
boundary elements.

\section{BASIC CODE STRUCTURE}
\label{sec:structures}

The boundary element library, \bemlib, has been written with a number
of goals in mind. The first of these is generality: users should be
able to solve different physical problems and implement different
solution techniques and elements as required. The second is that
existing wheels should be used rather than being reinvented: the codes
use other, high-quality, free libraries for various aspects of the
work. These are the GNU Triangulated Surface library (GTS) for
geometry handling~\cite{popinet00}, the GNU Scientific
Library~\cite{galassi-davies-etal05}, LAPACK~\cite{lapack99} in the
iterative solver and the GLIB library for portability functions and
special data structures, such as the tree used in the fast multipole
method. The GMSH suite of meshing and visualization
tools~\cite{geuzaine-remacle09} is used to generate geometries and
examine results. Using existing free software improves code
reliability and gives users an automatic upgrade as these codes are
improved.

The library and its programs have been written in~C, partly because~C
is nearly universally supported across the range of computing
platforms, partly because the other libraries used have~C interfaces,
and also because~C allows the use of programming techniques which make
extension and customization easier, for example, through the use of
loadable modules, an approach which is supported by the GLIB
library. The resulting code has been written as a number of separate
libraries to facilitate code re-use and to encourage generality in the
coding. 

\subsection{Problem definition}
\label{sec:problem}

For computational purposes, the partial differential equation for a
problem is defined by its Green's function. Boundary integral
equations have been applied in many areas of engineering and science
and here we concentrate on the Laplace and Helmholtz potential
problems which have applications in fluid dynamics~\cite{morino93} and
in acoustic scattering~\cite{pierce89,junger-feit93}. The form of the
integral equation is identical in each case, with only the Green's
function being different. For an unbounded domain containing
surface(s) $S$ with outward pointing normal $\mathbf{n}$, the
potential $\phi$ is given by:
\begin{subequations}
  \label{equ:pde}
  \begin{align}
    \label{equ:laplace}
    C(\mathbf{x})\phi(\mathbf{x}) &= \int_{S} 
    G\frac{\partial \phi}{\partial n_{1}}
    -
    \phi\frac{\partial G}{\partial n_{1}}\,\D S,
    \quad G = \frac{1}{4\pi R}\\
    \label{equ:helmholtz}
    C(\mathbf{x})\phi(\mathbf{x}) &= 
    \int_{S} 
    G\frac{\partial \phi}{\partial n_{1}}
    -
    \phi\frac{\partial G}{\partial n_{1}}\,\D S,
    \quad G = \frac{\E^{\J k R}}{4\pi R},
  \end{align}
\end{subequations}
where $k=\omega/c$ is the wavenumber of the Helmholtz problem,
$R=|\mathbf{x}-\mathbf{x}_{1}|$ and subscript $1$ denotes variables of
integration. The constant $C$ depends on the field point position
$\mathbf{x}$: 
\begin{align}
  \label{equ:def:C}
  C(\mathbf{x}) &=
  \left\{
    \begin{matrix}
      0 && \text{$\mathbf{x}$ inside $S$};\\
      1 && \text{$\mathbf{x}$ outside $S$};\\
      1 + \int_{S} \frac{\partial G_{0}}{\partial n_{1}}\,\D S &&
      \text{$\mathbf{x}$ on $S$}
    \end{matrix}
  \right.
\end{align}
with $G_{0}=1/4\pi R$. When $\mathbf{x}$ is on $S$ with the boundary
condition $\chi=\partial\phi/\partial n$ prescribed,
Equation~\ref{equ:pde} is a boundary integral equation for $\phi$ and
can be solved using the boundary element method.

The geometry $S$ is discretized into a number of elements with given
nodes $\mathbf{x}_{i}$, $i=1,\ldots,N$ and interpolation (shape)
functions $L_{i}$. This yields the system of equations:
\begin{align}
  \label{equ:matrix}
  \sum_{j=1}^{N} A_{ij}\phi_{j} &= \sum_{j=1}^{N} B_{ij}\chi_{j},
\end{align}
with
\begin{align*}
  A_{ij} &= C_{ij} + \sum_{m} \int L_{j} \frac{\partial G}{\partial
    n_{1}}\D S_{m},\\
  C_{ij} &= 
  \left\{
    \begin{matrix}
      1+\int \partial G_{0}/\partial n_{1}\,\D S, \quad i=j\\
      0, \quad i \neq j,
    \end{matrix}
  \right.\\
  B_{ij} &= \sum_{m} \int L_{j} G\, \D S_{m},
\end{align*}
where the summation over $m$ is taken on elements which contain
collocation point $j$, the shape function $L_{j}$ is the shape
function on element $m$ corresponding to point $j$ and $S_{m}$ is the
surface of element $m$. Conceptually, the boundary element method
consists of discretizing the surface, assembling the matrices $A$ and
$B$, and solving the system of Equation~\ref{equ:matrix}. In practice,
as we shall see below, this is not necessarily a feasible approach for
large problems, but it is the default method and one which can be used
as a baseline for assessing other techniques.

In \bemlib, the integrations in assembling the matrices are performed
using quadrature rules for singular integrands~\cite{khayat-wilton05},
Hayami's transformation for near-singular integrals~\cite{hayami05}
and the symmetric rules of Wandzura and
Xiao~\cite{wandzura-xiao03}. The system is solved using a library,
\sisl, based on LAPACK~\cite{lapack99}, which implements the
`templates' of Barrett et. al~\cite{barrett-berry-chan-demmel-etal94}.
The solver library allows for parallel solution of problems using the
MPI standard and also has a matrix-free option for use with
fast-multipole methods, as described below, \S\ref{sec:accelerated}. 

\subsection{Element types}
\label{sec:structures:elements}

Within the code of \bemlib, elements are represented as a collection
of triangles, based on the underlying GTS triangle data type. This
raises the question of how to link computational information to
geometric while retaining the freedom to introduce new element
types. Previous work using GTS surfaces~\cite{pantazopoulou06,%
  pantazopoulou-carley06,pantazopoulou-rice-carley05} introduced a new
data type for the vertices which allowed them to have an integer index
which could be used in assembling the system matrices. This is not a
satisfactory solution, however, when more general elements are to be
implemented. The first reason is that only linear triangular elements
based on the underlying triangles can be be used; the second is that
there is no good way to introduce discontinuities at sharp edges, an
essential part of representing general geometries.

\begin{figure}
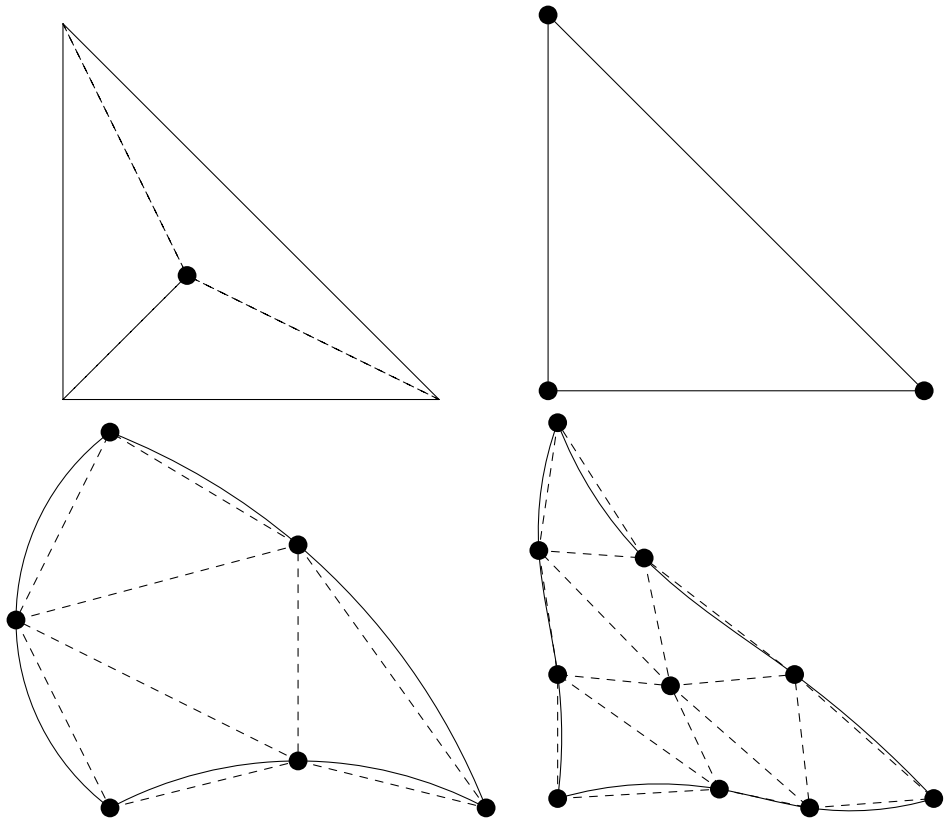

  \centering
  \begin{tabular}{cc}
  \includegraphics{ijnme09a-figs.6} &
  \includegraphics{ijnme09a-figs.3} \\
  \includegraphics{ijnme09a-figs.4} &
  \includegraphics{ijnme09a-figs.5}
  \end{tabular}
  \caption{Triangular element types: zero, first, second and third
    order: circles show computational nodes, solid lines the element
    boundary and dashed lines the GTS triangles.}
  \label{fig:elements}
\end{figure}

The solution which has been adopted is shown in
Figure~\ref{fig:elements}. Within the library, an element is composed
of a list of triangles, a list of collocation points with their
indices, a list of geometric nodes, interpolation functions for
surface data and interpolation functions for the geometry. For an
isoparametric element, the interpolation functions for the geometry
are the same as those for the surface data and the geometric nodes are
the same as the collocation points. This approach allows nodes to be
indexed independently on each element, to support discontinuities,
while allowing the underlying GTS surface to be geometrically valid.

Figure~\ref{fig:elements} shows the triangular elements which have
been implemented; quadrilateral elements have also been implemented
using the same approach. The shape functions employed are polynomials
and elements of order zero to three have been
developed. Figure~\ref{fig:elements} shows each element with its
underlying GTS triangles and the collocation points. In the case of a
zero order element, there is one collocation point, placed at the
centre of the triangular element. The element is made up of three GTS
triangles in order to allow for a constant solution across the
computational triangle. In the case of a linear element, the
computational triangle coincides with a GTS triangle and likewise the
collocation points. For the second and third order elements, the
element does not coincide with GTS triangles due to the element
curvature, with the underlying triangles being formed by joining the
collocation points with straight line segments.

Finally, a mesh is implemented as a GTS surface supplemented with a
list of elements, a lookup table connecting elements to triangles and
a lookup table connecting collocation points to their indices. These
lookup tables are implemented as hash tables, a well-known method for
efficient referencing of data.

\subsection{Detection and indexing of sharp edges}
\label{sec:edges}

\begin{figure}
  \centering
  \includegraphics{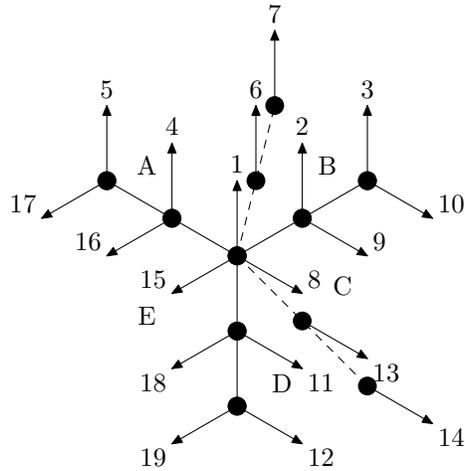}
  \caption{Renumbering the nodes at the corner of a cube}
  \label{fig:sharp:cube}
\end{figure}

In order to solve problems involving general geometries, a boundary
element code must have some method of dealing with sharp edges. In the
case of zero-order elements where collocation points lie on one, and
only one, element, there is no difficulty, but in the general case
where collocation points lie on more than one element, it is required
that the scheme be able to support a discontinuity in solution or
boundary condition at a sharp edge. As in previous work, this is
implemented by giving points of discontinuity multiple indices. This
allows for a discontinuity as the sharp edge is approached. An example
of such indexing, at the corner of a cube, is shown in
Figure~\ref{fig:sharp:cube}. Points are shown with their surface
normals and their indices, while elements are labelled with capital
letters. As an example of a sharp edge, nodes on the line between
elements~B and~C each have two indices while those between elements~B
and~A have only one: the surface is smooth at these points. The
exception is the corner of the cube where the node has three indices
corresponding to the three planes which meet there.

A method has been developed to automatically detect and index sharp
edges and corners, an important practical point in applications such
as aerodynamics~\cite{willis-peraire-white07} where the imposition of
an edge condition can be difficult and will ideally be performed
without user intervention. To index the nodes of a mesh, each element
is visited in turn. On each element, the currently unindexed nodes are
visited. If a node is shared with adjoining elements, the surface
normal at the node is computed on all of the elements. If the angle
between normals is less than some specified value, the surface is
taken to be smooth and the node is given one index. Otherwise, it is
given a different index for each normal which deviates from the
reference normal by more than the specified tolerance.

\subsection{Code structure}
\label{sec:structure:code}

\begin{figure}
  \centering
  \includegraphics{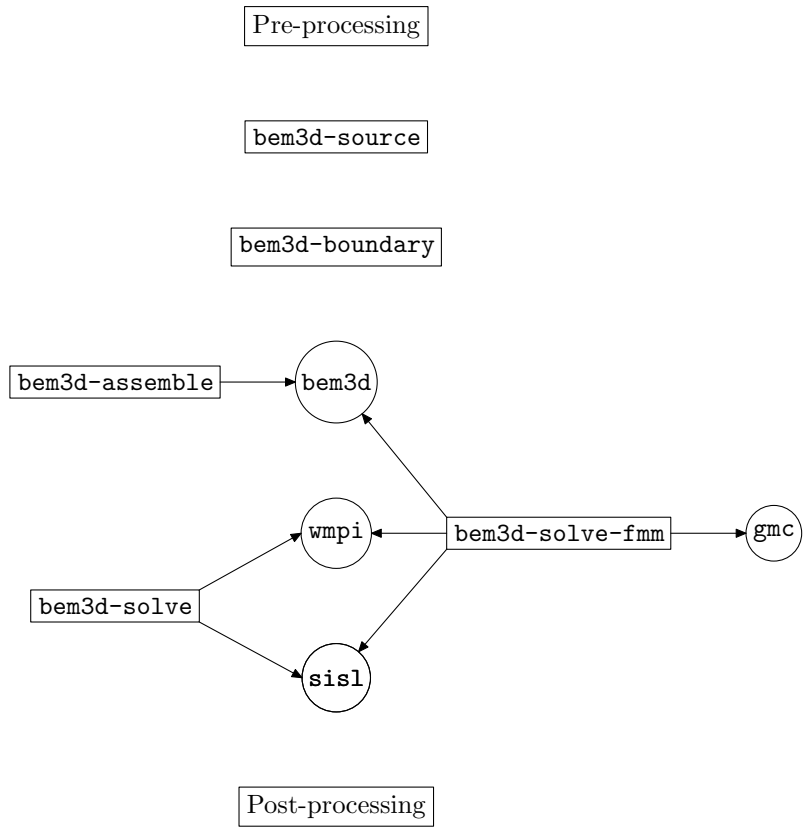}
  \caption{Structure of \bemlib\ codes: arrows indicate links to
    libraries. Executables are shown boxed and libraries circled.}
  \label{fig:structure}
\end{figure}

The structure of the codes in \bemlib\ is shown in
Figure~\ref{fig:structure}. The sequence shown contains all of the
steps which might be needed, but in many problems they will not all be
necessary. The first is a pre-processing stage to generate the
geometry and elements and to index the nodes. In most cases, the
geometry is generated using GMSH~\cite{geuzaine-remacle09} and
converted to the \bemlib\ format and indexed using a conversion
program. Boundary conditions are then supplied, either directly, or by
computing the field and its gradients from a prescribed source. If the
problem is being solved by direct solution of a matrix system, the
matrices are assembled and stored to be used by the solver; if the
fast multipole method is used, the information required for the matrix
multiplications is generated and applied directly in the same code. If
required, the field quantities can then be computed. Finally, any
post-processing and visualization which might be required are
performed by converting the solutions and mesh to the GMSH format.

The linkages between libraries and codes are shown in
Figure~\ref{fig:structure} with executables shown boxed and libraries
circled. The libraries which are used are the main \bemlib\ library,
\wmpi, a set of wrapper functions for interfacing to MPI, \sisl, a
simple iterative solver library and \gmc, which contains the code for
fast and accelerated multipole calculations. 

\section{ACCELERATED EVALUATION OF INTEGRALS}
\label{sec:accelerated}

Since its introduction, the fast multipole method has been used by a
number of workers to accelerate the boundary element and to reduce its
memory requirements~\cite{nishimura02}. The standard approach of
direct solution of the matrix system has time and memory requirements
which scale as $N^{2}$, which quickly exceed the resources available
on even the most powerful computers. The method implemented in
\bemlib\ allows problems to be solved directly on parallel systems
using the MPI standard but it was considered better to implement a
fast multipole method for solution of problems and for calculation of
radiated fields. The fast multipole method was originally developed
for point sources, rather than elements of finite extent, and is
usually implemented using spherical harmonics. In the method
implemented in \bemlib, the basic algorithm is similar to the original
adaptive fast multipole method~\cite{cheng-greengard-rokhlin99}, but
with two important differences. The first is the use of Taylor series
in place of spherical harmonics, as used by Tausch in his non-adaptive
method~\cite{tausch03,tausch04a}, in order to cater for a wide variety
of Green's functions. The second is a convergence radius criterion for
the near field, an idea developed~\cite{buchau-hafla-groh-rucker03} to
allow for finite size elements which may not be completely contained
within a box of the tree decomposing the computational surface. The
rest of this section details the sequence of procedures involved in
performing a fast multipole solution of the boundary element
equations.

\subsection{Hierarchical domain decomposition}
\label{sec:decomposition}

\begin{figure}
  \centering
  \includegraphics{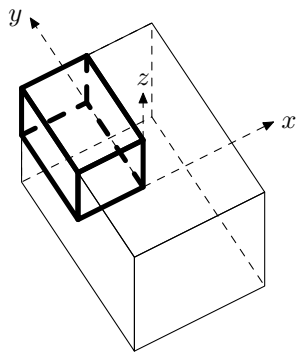}
  \caption{Decomposition of a box of points: the parent box is shown
    with one child box indicated in bold}
  \label{fig:boxtree}
\end{figure}

\begin{figure}
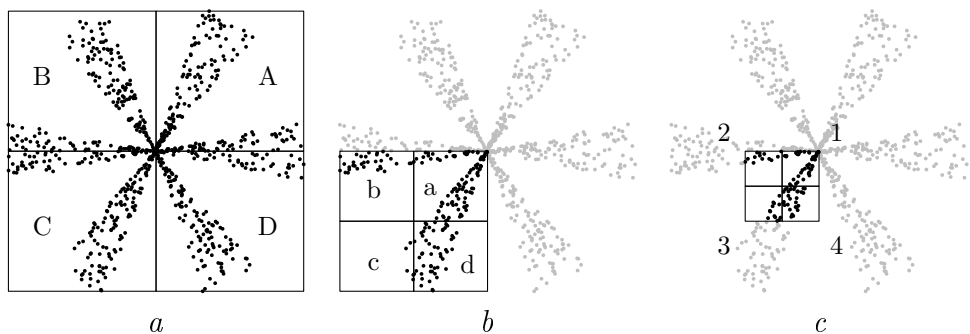

  \centering
  \begin{tabular}{ccc}
    \includegraphics{ijnme09a-figs.11} &
    \includegraphics{ijnme09a-figs.12} &
    \includegraphics{ijnme09a-figs.13} \\
    \textit{a} & \textit{b} & \textit{c} 
  \end{tabular}
  \caption{Decomposition of a set of points}
  \label{fig:decomposition}
\end{figure}

\begin{figure}
  \centering
  \includegraphics{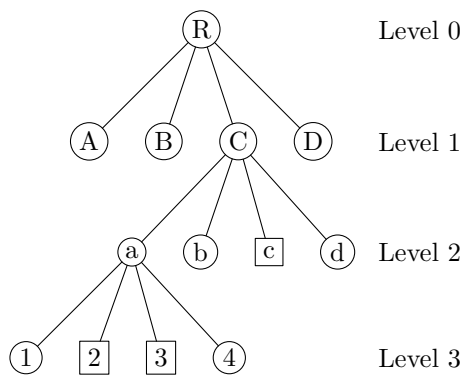}
  \caption{Tree structure for decomposition of
    Figure~\ref{fig:decomposition}}
  \label{fig:tree}
\end{figure}

The first step in a fast multipole method is to recursively decompose
the domain into a tree structure. The tree is made up of boxes aligned
with the coordinate axes with each box containing up to eight boxes,
formed by dividing the main box along each of the axes,
Figure~\ref{fig:boxtree}. The main box is called a `parent box' and
the boxes formed by subdivision are the `child boxes'. The
decomposition of the domain is initialized by forming a box which
contains all of the vertices of the computational surface $S$. This
box is said to be at level zero. The box is then subdivided to form
the level one boxes, which are themselves divided to form the level
two boxes and so on. At each stage of division, the number of vertices
in a box is checked. If it is less than some prescribed number $B$,
division of the box is terminated and the box is called a
`leaf'. Recursive division in this manner gives rise to a tree
structure where the box at each node of the tree encloses all of the
boxes in the nodes below it.

The process of subdivision is shown in two dimensions in
Figure~\ref{fig:decomposition}. An initial distribution of points is
enclosed by a level zero box which is divided into four level one
child boxes, labelled A, B, C and~D,
Figure~\ref{fig:decomposition}\textit{a}. Following the decomposition
of box~C, Figure~\ref{fig:decomposition}\textit{b} shows the level two
boxes a, b, c and~d. In this case, box~c has fewer than $B$ points and
is subdivided no more. Following the subdivision one more time, box~a
is divided into four level three boxes, 1, 2, 3 and~4, of which
boxes~2 and~3 are not divided again. 

Part of the resulting tree structure is shown in
Figure~\ref{fig:tree}, with leaf nodes of the tree shown boxed and
parent nodes shown circled. The parent node $R$ at level~0 has four
child nodes, corresponding to the four boxes of
Figure~\ref{fig:decomposition}\textit{a} and so forth. The tree
structure can be used to rapidly locate a point and as the basis for
fast integration methods. 

The final step in generating the tree is to attach the elements of the
mesh to the leaf nodes. The method adopted here is to list the
elements attached to the vertices of a leaf box and link this list to
the leaf node. 

\subsection{Taylor multipole expansions}
\label{sec:taylor}

A fast multipole method works by replacing a list of elements attached
to a box with a set of multipole coefficients which can be used to
compute the field due to the elements in the box far field, where `far
field' will be defined more precisely below. This was originally done
using an expansion of the field in spherical harmonics, but here
Taylor series expansions have been used since they allow the unified
treatment of a range of Green's functions~\cite{tausch03,tausch04a},
an important point in constructing a general code. 

The field integrals to be considered are of the form:
\begin{align}
  \label{equ:basic}
  I(\mathbf{x}) &= \int_{S_{b}} f(\mathbf{x_{1}})G(R)\,\D S_{b},
\end{align}
where $f(\cdot)$ is a source term and $S_{b}$ is the surface of the
elements inside some box of the tree decomposing the surface $S$. To
approximate $I(\mathbf{x})$, we expand the Green's function in a
Taylor series about some point $\mathbf{x}_{c}$ inside the box:
\begin{subequations}
  \begin{align}
    \label{equ:taylor:gfunc}
    G(R) & \approx \sum_{h=0}^{H}
    \sum_{m,n,k} 
    (-1)^{h} 
    G_{mnk}(x_{1}-x_{c})^{m} (y_{1}-y_{c})^{n}
    (z_{1}-z_{c})^{k},\\
    G_{mnk} &= 
    \frac{1}{m!}
    \frac{1}{n!}
    \frac{1}{k!}
    \left.
      \frac{\partial^{h}G}{\partial x^{m}\partial y^{n}\partial
        z^{k}}
      \right|_{\mathbf{x}_{1}=\mathbf{x}_{c}},
  \end{align}
\end{subequations}
where the summation over $m$, $n$ and $k$ is taken over all values of
$m+n+k=h$ and the symmetry relation $\partial G/\partial
x_{1}=-\partial G/\partial x$ has been used. The integral can then be
rewritten:
\begin{align}
  \label{equ:evaluation}
  f(\mathbf{x}) &\approx \sum_{h=0}^{H}
  \sum_{m,n,k} (-1)^{h}G_{mnk}I_{mnk},\\
  I_{mnk} &= \int_{S_{b}} f(\mathbf{x}_{1})(x_{1}-x_{c})^{m}
  (y_{1}-y_{c})^{n} (z_{1}-z_{c})^{k}\,\D S_{b}.\nonumber
\end{align}
The integrations can be performed the interpolation functions on each
element so that:
\begin{align}
  \label{equ:mpole:cft}
  I_{mnk} &= \sum_{i} f_{i}w^{(mnk)}_{i}
\end{align}
where the index $i$ runs over all points on elements connected to the
source box and $w_{i}^{(mnk)}$ is a weight precomputed as:
\begin{align}
  \label{equ:wt}
  w_{i}^{(mnk)} &= \int_{S_{b}} L_{i} (x_{1}-x_{c})^{m} (y_{1}-y_{c})^{n}
  (z_{1}-z_{c})^{k}\,\D S_{b}
\end{align}
which is independent of the underlying Green's function and so can be
stored and used in multiple problems.

To evaluate the field due to the source elements contained in a box,
the Taylor series can be used in an expansion about a point
$\mathbf{x}'_{c}$:
\begin{align*}
  f(\mathbf{x}) & \approx \sum_{l=0}^{L} \sum_{m,n,k} F_{mnk}
  (x-x'_{c})^{m}(y-y'_{c})^{n} (z-z'_{c})^{k}
\end{align*}
where the expansion coefficients $F_{mnk}$ are computed by
differentiation of Equation~\ref{equ:evaluation}:
\begin{align}
  \left.
    \frac{\partial^{q+r+s} f}{\partial x^{q}\partial y^{r} \partial
      z^{s}}
  \right|_{\mathbf{x}=\mathbf{x}_{c}'}
    & \approx
    \sum_{h=0}^{H}
    \sum_{m,n,k} (-1)^{h}G_{m+q,n+r,k+s}I_{mnk},\nonumber\\
    \label{equ:expansion}
    F_{qrs} &= \sum_{h=0}^{H}
    \sum_{m,n,k}
    (-1)^{h}\frac{1}{q!r!s!}G_{m+q,n+r,k+s}I_{mnk},
\end{align}
which can be used to generate a local expansion about the centre of
one box of the field due to another box.

Two basic tools are now required which allow the multipole expansions
of a tree of boxes to be used to accelerate the evaluation of the
field integrals. The first is the coefficient shift operator which
allows coefficients $I_{mnk}$ evaluated about one centre to be shifted
to another centre. This allows the coefficients about the centre of a
parent box to be evaluated from the coefficients of its child boxes
rather than having to be recomputed. Writing:
\begin{align*}
  I_{mnk}' &= \int_{S} f(\mathbf{x}_{1})(x_{1}-x_{c}')^{m}
  (y_{1}-y_{c}')^{n} (z_{1}-z_{c}')^{k}\,\D S,\\
  &= \int_{S} f(\mathbf{x}_{1})
  (x_{1}-x_{c} + (x_{c}-x_{c}'))^{m}
  (y_{1}-y_{c} + (y_{c}-y_{c}'))^{n}
  (z_{1}-z_{c} + (z_{c}-z_{c}'))^{k}\,\D S,
\end{align*}
and applying the binomial theorem:
\begin{align}
  \label{equ:mpole:shift}
  I_{mnk}' &= 
  \sum_{q=0}^{m} \sum_{r=0}^{n} \sum_{s=0}^{k}
  \binom{m}{q} \binom{n}{r} \binom{k}{s}
  (x_{c}-x_{c}')^{q} (y_{c}-y_{c}')^{r} (z_{c}-z_{c}')^{s}
  I_{(m-q)(n-r)(k-s)},
\end{align}
which allows the coefficients $I_{mnk}'$ about centre
$\mathbf{x}_{c}'$ to be computed in terms of coefficients $I_{mnk}$
about centre(s) $\mathbf{x}_{c}$. 

The second basic tool is the expansion shift operator which shifts the
expansion about one centre $\mathbf{x}_{c}$,
Equation~\ref{equ:expansion}, to give an expansion about another
centre $\mathbf{x}_{c}'$. Writing:
\begin{align}
  f(\mathbf{x}) &= 
  \sum_{l=0}^{L}\sum_{m,n,k}
  F'_{mnk} (x-x'_{0})^{m} (y-y'_{0})^{n} (z-z'_{0})^{k},\nonumber\\
  &=
  \sum_{l=0}^{L}\sum_{m,n,k} F_{mnk} 
  (x-x'_{0} + (x_{0}'-x_{0}))^{m} 
  (y-y'_{0} + (y_{0}'-y_{0}))^{n} 
  (z-z'_{0} + +(z_{0}'-z_{0}))^{k},\nonumber\\
  \label{equ:expansion:shift}
  &=
  \sum_{l=0}^{L}\sum_{m,n,k} F_{mnk} 
  \sum_{q=0}^{m}
  \sum_{r=0}^{n}
  \sum_{s=0}^{k}
  (x-x'_{0})^{m-q}(x_{0}'-x_{0})^{q} 
  (y-y'_{0})^{n-r}(y_{0}'-y_{0})^{r} 
  (z-z'_{0})^{k-s}(z_{0}'-z_{0})^{s},
\end{align}
from which the contribution of the original coefficients $F_{mnk}$ to
the shifted coefficients $F_{mnk}'$ can be identified. This allows a
parent box expansion to shifted to its children.

\subsection{Derivatives of Green's functions}
\label{sec:gfunc}

In order to compute the field due to a set of multipole coefficients,
we require the derivatives of the Green's function of a
problem. Tausch~\cite{tausch03} gives a recursive algorithm which
efficiently and stably generates the derivatives of a Green's function
up to some required order. 

Given a Green's function $G(R)$ which is a function of source-observer
distance $R=|\mathbf{r}|$, $\mathbf{r}=(x,y,z)$, the function:
\begin{align*}
  G^{(h)}(R) &= 
  \left(
    \frac{1}{R}\frac{\partial}{\partial R}
  \right)^{(h)}
  G(R),
\end{align*}
is defined. The sequence of functions $G^{(h)}$ can be computed
recursively. For the Laplace Green's function, $G=1/4\pi R$ and:
\begin{align*}
  G^{(0)}(R) &= \frac{1}{4\pi R},\quad G^{(h+1)} =
  -\frac{2h+1}{R^{2}}G^{(h)}(R),
\end{align*}
while for the Helmholtz Green's function, $G=\exp[\J k R]/4\pi R$:
\begin{align*}
  G^{(h+1)} &= -\frac{2h+1}{R^{2}}G^{(h)}(R)
  -\frac{k^{2}}{R^{2}}G^{(q-1)}(R),
\end{align*}
with the correction of a typographical error in the original
reference.

Given the sequence $G^{(h)}$, the derivatives of the functions with
respect to the components of $\mathbf{r}$ can be found from:
\begin{subequations}
  \label{equ:derivatives}
  \begin{align}
    \frac{\partial^{m+1+n+k} G^{(h)}}{\partial x^{m+1} \partial
      y^{m} \partial z^{k}}
    &= 
    m\frac{\partial^{m-1+n+k} G^{(h+1)}}{\partial x^{m-1} \partial
      y^{n} \partial z^{k}}
    +
    x\frac{\partial^{m+n+k} G^{(h+1)}}{\partial x^{m} \partial
      y^{m} \partial z^{k}},\\
    \frac{\partial^{m+n+1+k} G^{(h)}}{\partial x^{m} \partial
      y^{n+1} \partial z^{k}}
    &= 
    n\frac{\partial^{m+n-1+k} G^{(h+1)}}{\partial x^{m} \partial
      y^{n-1} \partial z^{k}}
    +
    y\frac{\partial^{m+n+k} G^{(h+1)}}{\partial x^{m} \partial
      y^{m} \partial z^{k}},\\
    \frac{\partial^{m+n+k+1} G^{(h)}}{\partial x^{m} \partial
      y^{n} \partial z^{k+1}}
    &= 
    k\frac{\partial^{m+n+k-1} G^{(h+1)}}{\partial x^{m} \partial
      y^{n} \partial z^{k-1}}
    +
    z\frac{\partial^{m+n+k} G^{(h+1)}}{\partial x^{m} \partial
      y^{n} \partial z^{k}}.
  \end{align}  
\end{subequations}
To compute the derivatives of the Green's function up to a given order
$H$, Tausch gives the following scheme~\cite{tausch03}:
\begin{enumerate}
\item compute $G^{(h)}$, $h=0,\ldots,H$;
\item compute $\partial^{m+n+k}G^{(h)}/\partial x^{m} \partial
  y^{n} \partial z^{k}$, for $m+n+k\leq H-h$, $h=H-1,\ldots,0$, using
  Equations~\ref{equ:derivatives};
\item set $\partial^{m+n+k}G/\partial x^{m} \partial y^{n} \partial
  z^{k}=\partial^{m+n+k}G^{(0)}/\partial x^{m} \partial y^{n} \partial
  z^{k}$, for $m+n+k=0,\ldots, H$. 
\end{enumerate}

\subsection{Box near field}
\label{sec:near:field}

\begin{figure}
  \centering
  \includegraphics{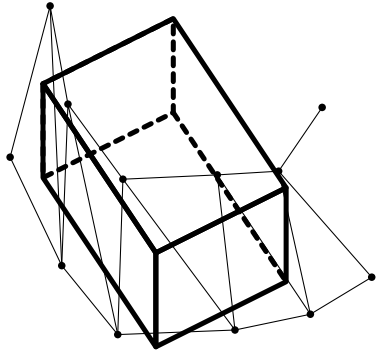}
  \caption{Elements stick out of their boxes}
  \label{fig:sticky:out}
\end{figure}

\begin{figure}
  \centering
  \includegraphics{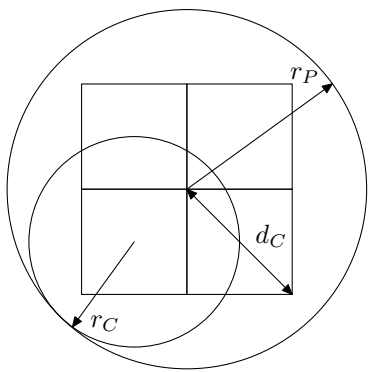}
  \caption{`Inheritance' of child convergence radius $r_{C}$ by parent
    box with convergence radius $r_{P}$} 
  \label{fig:radii}
\end{figure}

The tree decomposing the domain can be used to accelerate the
computation of field integrals, using the methods described in the
next two sections. The basic approach is to use the multipole
expansion about a box centre to compute the field `far' from the box
and to use full integration over the box elements `close' to the
box. This requires a definition of `near' and `far' which can be used
in deciding which procedure to use. In the original version of the
fast multipole method~\cite{cheng-greengard-rokhlin99}, the point
sources used were completely contained in a box and the near field of
a box was defined as its neighbours. When elements of finite extent
are used, they will inevitably stick out of the boxes to which they
are attached and it is not clear how to define the boundary of the
`far field' of the box. 

A number of approaches have been discussed to deal with this problem,
some of them limited to zero order elements. Tausch~\cite{tausch04a}
defines a separation ratio which can be used in deciding when boxes
lie in each others' near field. This has the disadvantage that it
requires the boxes to be at the same level in the tree, i.e. that they
be the same size. Another approach, employed in a spherical harmonic
fast multipole code~\cite{buchau-hafla-groh-rucker03}, is to use the
convergence radius of the expansion about a box centre define the box
near field. This approach has been used in \bemlib, with the
convergence radius $r_{C}$ given by the maximum distance between the
box centre and a vertex of an element attached to the box (which may
well not lie in the box proper). A point lies in the far field of the
box if its distance to the box centre is greater than $r_{C}$. Two
boxes, which need not be the same size, lie in each others' far fields
if the distance between their centres is greater than the sum of their
convergence radii.

Finally, the convergence radius $r_{P}$ of a parent box can be derived
from the radii of its child boxes using the approach shown in
Figure~\ref{fig:radii}:
\begin{align}
  \label{equ:radii}
  r_{P} &= r_{C} + d_{C}/2,
\end{align}
where $r_{C}$ is the maximum of the convergence radii of the child
boxes and $d_{C}$ is the length of a child box's diagonal.

\subsection{Accelerated field evaluation}
\label{sec:amm}

The first principal application of the basic techniques of the
previous sections is in accelerated evaluation of the field due to a
known source distribution over a surface. Given the tree decomposition
of Section~\ref{sec:decomposition}, the multipole moments computed at
each leaf node and shifted upwards in the tree,
Section~\ref{sec:taylor} and a field point $\mathbf{x}$, the field may
be evaluated by traversing the tree and computing the contribution
from each box of the tree in turn, descending the tree only far enough
to evaluate the contribution of a branch to sufficient accuracy. 

The algorithm may be summarized as follows:
\begin{enumerate}
\item set $I(\mathbf{x})=0$;
\item set the current box to the root of the tree at level~0;
\item \label{amm:loop} for the current box:
  \begin{enumerate}
  \item if the distance to the box centre
    $|\mathbf{x}-\mathbf{x}_{c}|>r_{C}$, evaluate the multipole
    expansion of Equation~\ref{equ:expansion} and add to
    $I(\mathbf{x})$; terminate descent of this branch;
  \item if the distance to the box centre
    $|\mathbf{x}-\mathbf{x}_{c}|\leq r_{C}$ and the current box is a
    parent, repeat step~\ref{amm:loop} for each child box;
  \item if the distance to the box centre
    $|\mathbf{x}-\mathbf{x}_{c}|\leq r_{C}$ and the current box is a
    leaf node, evaluate the field by direct integration over the
    elements of the box and add to $I(\mathbf{x})$; terminate descent
    of this branch.
  \end{enumerate}
\end{enumerate}
This algorithm can be used to evaluate the field at a small number of
points which may be far from the surface $S$: it is efficient and
avoids the setup costs involved in the fast multipole method
proper. When the field must be evaluated at a large number of points,
for matrix multiplication, say, the fast multipole method described in
the next section is used.

\subsection{Fast matrix multiplication}
\label{sec:fmm}

In order to further accelerate the computation of integrals in the
boundary element method, the technique of the previous section can be
extended to a true fast multipole method, at the expense of some extra
pre-processing. This yields a method which speeds up the matrix
multiplications required for the solution of the boundary element
equations by using local expansions of the field about the centre of
each leaf node of the tree.

\begin{figure}
  \centering
  \includegraphics{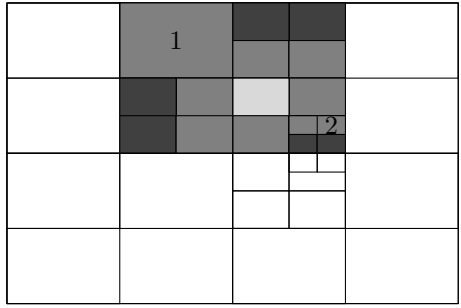}
  \caption{Interaction and near field lists for a tree box}
  \label{fig:lists}
\end{figure}

The algorithm is similar to that of the previous section but with some
extra procedures related to the evaluation of the near field terms and
the separation of boxes into near and far field. The first step is, as
before, to generate the tree decomposing the surface, including the
multipole coefficients and convergence radii for the boxes. Each box
now has linked to it two lists: the near-field list and the
interaction list. The near-field list consists of boxes whose
contribution to the field in the box must be computed by direct
integration over the attached elements. The interaction list is made
up of boxes whose contribution can be computed using the multipole
expansion. Figure~\ref{fig:lists} shows an example. The light grey box
in the middle of the grid is the box whose lists are shown. The darker
boxes are those in the near-field list, using the criterion of
Section~\ref{sec:near:field} which states that boxes are
well-separated iff:
\begin{align}
  \label{equ:separated}
  |\mathbf{x}_{c}^{(1)}-\mathbf{x}_{c}^{(2)}| > r_{C}^{(1)} +
  r_{C}^{(2)},
\end{align}
where $\mathbf{x}_{c}^{(i)}$ and $r_{c}^{(i)}$ are the centre and
radius of convergence of each of the boxes. The darkest boxes are
those on the interaction list whose contribution to the field in the
main box can be computed using multipole expansions. The field in the
box is computed as a sum of contributions from the near field and
interaction list boxes and from the local expansion about the centre
of the parent box which is computed in the same way. In
Figure~\ref{fig:lists}, the box labelled 1 is in the near field list,
even though it is larger than the main box, because it is a leaf node
and must contribute directly to the field. The box labelled 2 is in
the near field list, even though it does not touch the main box,
because it does not meet the separation criterion of
Equation~\ref{equ:separated}.

The near-field and interaction lists can be generated using the
following algorithm, applied to each box $B$ descending the tree
starting at level zero:
\begin{enumerate}
\item initialize the interaction and near-field lists of $B$ to be
  empty;
\item traverse the near field list of the box's parent $P$ and for
  each box $N$ in the list:
  \begin{enumerate}
  \item if $N$ is a leaf node:
    \begin{enumerate}
    \item if $N$ and $B$ are well-separated, add $N$ to the
      interaction list of $B$;
    \item if $N$ and $B$ are not well-separated, add $N$ to the
      near-field list of $B$;
    \end{enumerate}
  \item if $N$ is not a leaf node, traverse the child boxes $C$ of $N$
    and:
    \begin{enumerate}
    \item if $C$ and $B$ are well-separated, add $C$ to the
      interaction list of $B$;
    \item if $C$ and $B$ are not well-separated, add $C$ to the
      near-field list of $B$;
    \end{enumerate}
  \end{enumerate}
\end{enumerate}

Given a tree with multipole coefficient weights $w^{(mnk)}$ up to
order $m+n+k=H$, and near-field and interaction lists for each box, a
matrix multiplication can be performed as follows:
\begin{enumerate}
\item for each leaf box, compute the multipole moments $I_{mnk} =
  \sum_{i} f_{i}w^{(mnk)}_{i}$, $m+n+k \leq H$;
\item traverse tree from bottom to top, calculating box moments by
  accumulating the contribution from child boxes using
  Equation~\ref{equ:mpole:shift};
\item traverse tree from top to bottom, computing local expansions to
  order $L$:
  \begin{enumerate}
  \item shift parent local expansion to box centre using
    Equation~\ref{equ:expansion:shift};
  \item add local expansion terms due to boxes in interaction list;
  \end{enumerate}
  each leaf box now has an expansion about its centre which gives the
  field in the box due to all other boxes except those in the
  near-field list;
\item for each leaf box, evaluate the local expansion at each enclosed
  vertex and add the contribution from boxes in the near-field list.
\end{enumerate}
In practice, the near-field contribution is precomputed and stored as
a sparse matrix for re-use in the solution procedure and the expansion
order $L$ is a function of depth in the tree~\cite{tausch04a} in order
to avoid excessively high order expansions in small boxes. The order
$L_{l}$ at level $l$ is set to
$L_{l}=\min(L_{\max},L_{\min}+l_{\max}-l)$.

\section{NUMERICAL TESTS}
\label{sec:tests}

The performance of the boundary element codes has been assessed in
terms of accuracy, speed and memory. The test uses a point source
placed inside the surface to generate a potential and a boundary
condition (potential gradient). The system is solved for the boundary
condition which should recover the original surface potential due to
the point source. The error measure is the rms difference between the
original and computed potential. Two geometries have been used, a unit
sphere and a cube of edge length~2 centred on the origin. The sphere
is generated by successive refinement of an initial surface, allowing
tests to be carried out on a smooth surface with control over the
number of vertices. The cube is generated using
GMSH~\cite{geuzaine-remacle09} with successively smaller edge
lengths. This tests the ability of \bemlib\ to deal with sharp edges
using the methods outlined earlier. Tests have been conducted with
first and second order elements and the Laplace and Helmholtz ($k=2$)
equations have been solved. The fast multipole method was used to
solve problems of all sizes and the direct matrix method was used for
the smaller problems which would fit in memory. Calculations were
performed on a~3\giga\hertz\ Pentium~4 with 1Gb of memory.

Data reported are the number of vertices, the maximum number of
vertices $B$ in a tree box, the setup time and solution time and the
maximum memory used during solution. For matrix solution, the setup
time is the time required to assemble the system matrices. For the
fast multipole method, the setup time is that needed to generate the
tree for the mesh and compute and store the near-field matrix. In
neither case is the time needed for disk storage and recovery
included. In the fast multipole case, the time reported is the total
setup time for one problem. In practice, because the multipole
coefficient weights are independent of the problem being solved, the
setup time for multipole problems could be reduced by re-using the
weights. In both cases, the solution time reported is that needed for
solution using the stabilized biconjugate gradient
method~\cite{barrett-berry-chan-demmel-etal94}.

As a reference case, the fast multipole solver uses a minimum
expansion order of~3 and a maximum of~8. For larger problems, the
error ceases to reduce with element size. This appears to be due to
the multipole precision being larger than the discretization
error. As a check, a number of the larger problems were solved with
the minimum expansion order increased to~4. These results are included
in the tabulated data and indicated by an asterisk.

\begin{table}
\caption{Code performance for Laplace equation on sphere using
first  order elements}
\label{tab:sphere-L:1}
\begin{center}
    \begin{small}
\begin{tabular}{lrrrrrrrrr}
\toprule
& \multicolumn{5}{c}{FMM}& \multicolumn{4}{c}{Direct} \\
$N$ & $B$ & $\epsilon$ & $t_{\text{setup}}/\second$ & $t/\second$ & Mb & $\epsilon$ & $t_{\text{setup}}/\second$ & $t/\second$ & Mb\\
\midrule
 42 & 8 & $3.03\times10^{-2}$ & 1 & 0 & 2 &  $5.31\times10^{-3}$ & 1 & 0 & 0 \\
 162 & 16 & $1.55\times10^{-2}$ & 6 & 3 & 5 &  $1.38\times10^{-3}$ & 5 & 0 & 2 \\
 642 & 32 & $3.48\times10^{-4}$ & 40 & 7 & 10 &  $3.46\times10^{-4}$ & 80 & 0 & 8 \\
 2562 & 64 & $9.03\times10^{-5}$ & 473 & 8 & 49 &  $8.62\times10^{-5}$
 & 1322 & 0 & 104 \\
2562$^{*}$ & 64 & $8.59\times10^{-5}$ & 491 & 11 & 49 \\
 10242 & 128 & $7.32\times10^{-5}$ & 1758 & 54 & 203 & \\
10242$^{*}$ & 128 & $2.96\times10^{-5}$ & 1897 & 70 & 203 \\
\bottomrule
\end{tabular}
    \end{small}
  \end{center}
 \end{table}
\begin{table}
\caption{Code performance for Laplace equation on sphere using
second order elements}
\label{tab:sphere-L:2}
\begin{center}
    \begin{small}
\begin{tabular}{lrrrrrrrrr}
\toprule
& \multicolumn{5}{c}{FMM}& \multicolumn{4}{c}{Direct} \\
$N$ & $B$ & $\epsilon$ & $t_{\text{setup}}/\second$ & $t/\second$ & Mb & $\epsilon$ & $t_{\text{setup}}/\second$ & $t/\second$ & Mb\\
\midrule
 162 & 8 & $2.04\times10^{-4}$ & 5 & 3 & 5 &  $2.03\times10^{-4}$ & 3 & 0 & 2 \\
 642 & 16 & $2.55\times10^{-5}$ & 25 & 7 & 11 &  $2.20\times10^{-5}$ & 32 & 0 & 8 \\
 2562 & 32 & $3.33\times10^{-5}$ & 102 & 47 & 37 &  $2.32\times10^{-6}$ & 474 & 0 & 104 \\
 10242 & 64 & $4.82\times10^{-5}$ & 749 & 53 & 193 & \\
 40962 & 128 & $8.12\times10^{-5}$ & 3611 & 264 & 842 & \\
\bottomrule
\end{tabular}
    \end{small}
  \end{center}
 \end{table}

\begin{table}
\caption{Code performance for Helmholtz equation on sphere using
first  order elements}
\label{tab:sphere-H:1}
\begin{center}
    \begin{small}
\begin{tabular}{lrrrrrrrrr}
\toprule
& \multicolumn{5}{c}{FMM}& \multicolumn{4}{c}{Direct} \\
$N$ & $B$ & $\epsilon$ & $t_{\text{setup}}/\second$ & $t/\second$ & Mb & $\epsilon$ & $t_{\text{setup}}/\second$ & $t/\second$ & Mb\\
\midrule
 42 & 8 & $3.46\times10^{-2}$ & 1 & 0 & 2 &  $5.14\times10^{-3}$ & 1 & 0 & 430 \\
 162 & 16 & $1.45\times10^{-2}$ & 10 & 20 & 6 &  $1.28\times10^{-3}$ & 6 & 0 & 3 \\
 642 & 32 & $3.30\times10^{-4}$ & 68 & 35 & 12 &  $3.15\times10^{-4}$ & 118 & 0 & 15 \\
 2562 & 64 & $1.18\times10^{-4}$ & 843 & 40 & 70 &
 $7.73\times10^{-5}$ & 1758 & 1 & 207 \\
 2562$^{*}$ & 64 & $7.75\times10^{-5}$ & 1398 & 25 & 122 \\
 10242 & 128 & $1.83\times10^{-4}$ & 3020 & 284 & 284 & \\
 10242$^{*}$ & 128 & $2.06\times10^{-5}$ & 4721 & 231 & 475 & \\
\bottomrule
\end{tabular}
    \end{small}
  \end{center}
 \end{table}
\begin{table}
\caption{Code performance for Helmholtz equation on sphere using
second order elements}
\label{tab:sphere-H:2}
\begin{center}
    \begin{small}
\begin{tabular}{lrrrrrrrrr}
  \toprule
  & \multicolumn{5}{c}{FMM}& \multicolumn{4}{c}{Direct} \\
  $N$ & $B$ & $\epsilon$ & $t_{\text{setup}}/\second$ & $t/\second$ & Mb & $\epsilon$ & $t_{\text{setup}}/\second$ & $t/\second$ & Mb\\
  \midrule
  162 & 16 & $1.71\times10^{-2}$ & 5 & 9 & 6 &  $3.62\times10^{-4}$ &
  3 & 0 & 0 \\ 
  642 & 32 & $8.00\times10^{-5}$ & 31 & 23 & 14 &  $4.30\times10^{-5}$
  & 39 & 0 & 15 \\ 
  2562 & 64 & $9.64\times10^{-5}$ & 247 & 25 & 74 &
  $4.64\times10^{-6}$ & 583 & 1 & 206 \\ 
  2562$^{*}$ & 64 & $7.75\times10^{-5}$ & 1398 & 25 & 122\\
  10242 & 128 & $1.38\times10^{-4}$ & 888 & 165 & 282 & \\
  10242$^{*}$ & 128 & $2.06\times10^{-5}$ & 4721 & 231 & 475 & \\
  \bottomrule
\end{tabular}
    \end{small}
  \end{center}
 \end{table}

\begin{table}
\caption{Code performance for Laplace equation on cube using
first  order elements}
\label{tab:cube-L:1}
  \begin{center}
    \begin{small}
\begin{tabular}{lrrrrrrrrr}
\toprule
& \multicolumn{5}{c}{FMM}& \multicolumn{4}{c}{Direct} \\
$N$ & $B$ & $\epsilon$ & $t_{\text{setup}}/\second$ & $t/\second$ & Mb & $\epsilon$ & $t_{\text{setup}}/\second$ & $t/\second$ & Mb\\
\midrule
 117 & 5 & $2.22\times10^{-3}$ & 7 & 9 & 5 &  $2.22\times10^{-3}$ & 4 & 0 & 0 \\
 431 & 10 & $4.85\times10^{-4}$ & 21 & 21 & 9 &  $4.87\times10^{-4}$ & 30 & 0 & 5 \\
 1763 & 20 & $1.17\times10^{-4}$ & 114 & 81 & 29 &
 $1.19\times10^{-4}$ & 552 & 0 & 50 \\
 7281 & 40 & $2.55\times10^{-5}$ & 618 & 124 & 98 & \\
 28029 & 80 & $2.70\times10^{-5}$ & 5003 & 227 & 512 & \\
 28029$^{*}$ & 80 & $7.85\times10^{-6}$ & 17691 & 391 & 854 & \\
\bottomrule
\end{tabular}
    \end{small}
  \end{center}
 \end{table}
\begin{table}
\caption{Code performance for Laplace equation on cube using
second order elements}
\label{tab:cube-L:2}
  \begin{center}
    \begin{small}
\begin{tabular}{lrrrrrrrrr}
\toprule
& \multicolumn{5}{c}{FMM}& \multicolumn{4}{c}{Direct} \\
$N$ & $B$ & $\epsilon$ & $t_{\text{setup}}/\second$ & $t/\second$ & Mb & $\epsilon$ & $t_{\text{setup}}/\second$ & $t/\second$ & Mb\\
\midrule
 378 & 10 & $1.25\times10^{-4}$ & 27 & 47 & 8 &  $1.24\times10^{-4}$ & 10 & 0 & 4 \\
 1538 & 20 & $1.71\times10^{-5}$ & 69 & 69 & 27 &  $5.61\times10^{-6}$ & 154 & 0 & 39 \\
 1538$^{*}$ & 20 & $6.90\times10^{-6}$ & 73 & 94 & 27\\
 6674 & 40 & $9.10\times10^{-6}$ & 361 & 147 & 106 &
 $2.78\times10^{-7}$ & 2982 & 1 & 695 \\
 6674$^{*}$ & 40 & $5.45\times10^{-7}$ & 614 & 279 & 147 \\
 28362 & 80 & $2.09\times10^{-5}$ & 2551 & 253 & 636 & \\
\bottomrule
\end{tabular}
    \end{small}
  \end{center}
 \end{table}

\begin{table}
\caption{Code performance for Helmholtz equation on cube using
first  order elements}
\label{tab:cube-H:1}
  \begin{center}
    \begin{small}
\begin{tabular}{lrrrrrrrrr}
\toprule
& \multicolumn{5}{c}{FMM}& \multicolumn{4}{c}{Direct} \\
$N$ & $B$ & $\epsilon$ & $t_{\text{setup}}/\second$ & $t/\second$ & Mb & $\epsilon$ & $t_{\text{setup}}/\second$ & $t/\second$ & Mb\\
\midrule
 117 & 5 & $3.46\times10^{-3}$ & 8 & 67 & 5 &  $3.47\times10^{-3}$ & 5 & 0 & 2 \\
 431 & 10 & $7.34\times10^{-4}$ & 46 & 158 & 34 &  $7.43\times10^{-4}$ & 37 & 0 & 8 \\
 1763 & 20 & $1.69\times10^{-4}$ & 269 & 489 & 36 &  $1.68\times10^{-4}$ & 657 & 1 & 98 \\
 7281 & 40 & $5.11\times10^{-5}$ & 854 & 408 & 128 & \\
 28029 & 80 & $7.18\times10^{-5}$ & 6246 & 1142 & 713 & \\
\bottomrule
\end{tabular}
    \end{small}
  \end{center}
 \end{table}
\begin{table}
\caption{Code performance for Helmholtz equation on cube using
second order elements}
\label{tab:cube-H:2}
  \begin{center}
    \begin{small}
\begin{tabular}{lrrrrrrrrr}
\toprule
& \multicolumn{5}{c}{FMM}& \multicolumn{4}{c}{Direct} \\
$N$ & $B$ & $\epsilon$ & $t_{\text{setup}}/\second$ & $t/\second$ & Mb & $\epsilon$ & $t_{\text{setup}}/\second$ & $t/\second$ & Mb\\
\midrule
 378 & 10 & $2.49\times10^{-4}$ & 31 & 168 & 10 &  $2.45\times10^{-4}$ & 12 & 0 & 6 \\
 1538 & 20 & $5.73\times10^{-5}$ & 81 & 313 & 34 &
 $9.12\times10^{-6}$ & 189 & 1 & 76 \\
 1538$^{*}$ & 20 & $1.66\times10^{-5}$ & 87 & 421 & 34 \\
 6674 & 40 & $2.51\times10^{-5}$ & 431 & 1149 & 147 & \\
 6674$^{*}$ & 40 & $1.35\times10^{-6}$ & 775 & 1672 & 205 \\
 28362 & 80 & $5.36\times10^{-5}$ & 3009 & 901 & 892 & \\
\bottomrule
\end{tabular}
    \end{small}
  \end{center}
 \end{table}

The first data presented are for the Laplace equation solved on a
sphere, Tables~\ref{tab:sphere-L:1} and~\ref{tab:sphere-L:2}. The
number of points increases by a factor of about four between test
cases and the computing demands increase at the expected rate, with
the memory requirement scaling roughly as $N^{2}$ for the matrix
solution and roughly linearly for the fast multipole method. This is
also true for the setup time but not for the solution time, which for
the matrix method is less than one second in all cases while it
increases roughly linearly for the fast multipole method. For the
smaller problems $N\leq2562$, the error is essentially the same for
both solution methods but beyond this point the error from the fast
multipole method stops falling. This behaviour can be cured by
increasing the minimum expansion order as shown by the test cases
marked with an asterisk in Table~\ref{tab:sphere-L:1} where the
reduction in error with $N$ has been restored. It can also be noted
that, even without re-using multipole weights, the setup time for the
fast multipole method is much less than that for the direct solver,
far outweighing any advantage in solution time for the direct method,
even for relatively small problems.

The solvers behave similarly when applied to the Helmholtz
problem. The decline in error with $N$ is similar, with a need to
increase the minimum expansion order at large $N\geq2562$. 

The second test case for the Laplace and Helmholtz solvers is that of
a cube of edge length two, used to check the performance of the solver
on a geometry with sharp edges, a type of problem known to be quite
ill-conditioned. In the case of the Laplace equation,
Tables~\ref{tab:cube-L:1} and~\ref{tab:cube-L:2}, the fast multipole
solver is comparable in accuracy to the direct method, with the caveat
that the minimum expansion order must be increased for the larger
problems, but is far superior in time and memory
consumption. Likewise, the results for the Helmholtz equation show the
expected reduction in error with vertex number $N$ and a much lower
setup time for the fast multipole method, although the solution time
is rather large for small $N$ using second order elements. This
appears to be due to the inherent ill-conditioning of the problem and
because the ratio of edge vertices to non-edge vertices is greater for
smaller problems. No attempt was made to optimize the mesh to improve
the handling of sharp edges, as in other studies~\cite{tausch04a},
since it is intended that \bemlib\ be able to handle unstructured
meshes produced by a standard mesh generator, without requiring
intervention from the user.

\section{CONCLUSIONS}
\label{sec:conclusions}

A new library has been developed for the boundary element solution of
general problems. The library is free software released under the GNU
General Public Licence. It incorporates a new adaptive fast multipole
method for boundary element problems which gives comparable accuracy
to direct matrix solution at a much lower cost in terms of time and
memory, making possible the solution of problems which would otherwise
be too large to solve with the available resources. It includes higher
order elements and has provision for the addition of user-defined
elements and Green's functions without extensive recoding.

Development of \bemlib\ continues with the intention of adding further
element types and a parallel implementation of the fast multipole
method. 

\bibliography{abbrev,maths,misc,vortex,scattering,aerodynamics}

\end{document}